\documentclass[aps,prl,twocolumn,superscriptaddress,showpacs]{revtex4-1}
\usepackage{graphicx}
\usepackage{dcolumn}
\usepackage{bm}
\usepackage{subfigure}
\usepackage{amsmath}
\usepackage{hyperref}
\usepackage{xcolor}

\newcommand{\ra}{\rightarrow}

\newcommand{\bs}{\boldsymbol}

\newcommand{\SRO}{Sr$_2$RuO$_4$}

\usepackage{times}

\begin{document}
\title{Possible 3D nematic odd-parity pairing in \SRO: experimental evidences and predictions}
\author{Wen Huang}
\affiliation{Institute for Advanced Study, Tsinghua University, Beijing 100084, China}
\author{Yi Zhou}
\email{yizhou@zju.edu.cn}
\affiliation{Department of Physics, Zhejiang University, Hangzhou, Zhejiang 310027, China}
\affiliation{CAS Center for Excellence in Topological Quantum Computation, University of Chinese Academy of Sciences, Beijing 100190, China}
\author{Hong Yao}
\email{yaohong@tsinghua.edu.cn}
\affiliation{Institute for Advanced Study, Tsinghua University, Beijing 100084, China}
\affiliation{ State Key Laboratory of Low Dimensional Quantum Physics, Tsinghua University, Beijing 100084, China }
\date{\today}

\begin{abstract}
Due to the presence of a nontrivial three-dimensional spin-orbital entanglement, \SRO~may be a time-reversal invariant nematic $p$-wave superconductor with coexisting in-plane and out-of-plane pairings.
Here we discuss various signatures of such a state if the out-of-plane pairing component is dominant. First, the enhancement of the superconducting $T_c$ under in-plane uniaxial strains is nearly a quadratic function of the strain, because the out-of-plane pairing lacks a linear-order coupling to the strain. Second, when the strain applies along a certain in-plane direction, the nematic $p$-wave pairing exhibits only a single phase transition as the temperature is lowered. These are consistent with several recent uniaxial strain measurements, which are otherwise hard to reconcile with chiral $p$-wave order. We further show that the nematic $p$-wave state can be distinguished from the chiral $p$-wave state through the velocity jumps of certain sound waves at the onset of superconductivity. Possible implications for $\mu$SR experiment under strain and NMR Knight shift measurement are also discussed. 
\end{abstract}

\maketitle
{\bf Introduction:} Nearly one quarter century has past since superconductivity was discovered in \SRO~\cite{Maeno:94}, its putative unconventional pairing symmetry has been drawing increasing attention \cite{Maeno:01,Mackenzie:03,Kallin:09,Kallin:12,Maeno:12,Liu:15,Kallin:16,Mackenzie:17}. Of particular interest is the prospect of it being a topological chiral $p$-wave superconductor. Such a superconductor allows half-quantum superconducting vortices that host Majorana zero modes -- which obey non-Abelian braiding statistics \cite{Goore:91,Read:00,Volovik:99,Ivanov:01}. Hence this material is a promising candidate platform for topological quantum computation \cite{Kitaev:03,Nayak:08}. Multiple experimental signatures consistent with the chiral $p$-wave superconducting order, including time-reversal symmetry breaking (TRSB), spin-triplet, and odd-parity pairing \cite{Luke:98,Ishida:98,Duffy:00,Nelson:04,Xia:06}, have indeed been reported early on. However, a number of experimental observations still stand at odds with chiral $p$-wave \cite{Kallin:12,Maeno:12,Mackenzie:17}. In the following, we enumerate a few that relate most closely to the proposal of nematic $p$-wave pairing \cite{Huang:18}.

First, the predicted spontaneous surface current has not thus far been definitively detected \cite{Matsumoto:99,Furusaki:01,Kirtley:07,Hicks:10,Curran:14}. This has led to a number of explanations, either within \cite{Ashby:09,Sauls:11,Imai1213,Lederer:14,Bouhon:14,Huang:15,Scaffidi:15,Zhang:17a,Etter:17} or outside \cite{Huang:14,Tada:15} the chiral $p$-wave framework. In the former, the non-topological nature of edge current implies that, unless the $p$-wave pairing coincides with some fine-tuned anisotropic form, the current is ought to be nonvanishing at a pristine sample surface \cite{Huang:15,Volovik:14}. Second, in the presence of symmetry-lowering perturbations, such as uniaxial in-plane strain and in-plane magnetic field, the degeneracy of the two chiral components shall be lifted \cite{Sigrist:91}. This foretells two consecutive phase transitions as the temperature is lowered. However, until now there has been no compelling evidence for the split phase transitions \cite{Mao:00,Yonezawa:14}. Third, the aforementioned degeneracy lifting is expected to exhibit a linear cusp between tensile and compressive strains. However,   the observed $T_c$ enhancement follows a nonlinear line shape \cite{Hicks:14,Steppke:17}, resembling instead a quadratic function according to a more recent scanning SQUID measurement \cite{Watson:18}.

In regards to the superconducting mechanism, many early experiments were interpreted on the basis of the dominant-$\gamma$-band scenario, in light of the proximity to van-Hove singularity of this band \cite{Mackenzie:03,Agterberg:97}. However, recent theoretical analyses have reached diverse conclusions \cite{Raghu:10,Huo:13,Wang:13,Scaffidi:14,Tsuchiizu:15,Huang:16,Zhang:17b,Liu:17,Wang:18,Gingras:18}. In particular, it was suggested by some that the sizable SOC between the Ru $t_{2g}$-orbitals, of the order $\eta_\text{soc}/E_F \sim 0.1$ \cite{Haverkort:08,Veenstra:14,Fatuzzo:15}, induces noticeable interband Cooper pair scattering and could eventually induce comparable gaps on all three bands \cite{Scaffidi:14,Tsuchiizu:15,Huang:16,Zhang:17b,Wang:18}.

SOC has other profound consequences. When it is accompanied by interlayer inter-orbital hopping between the $t_{2g}$-orbitals, the electronic structure acquires a nontrivial three-dimensional (3D) spin-orbital entanglement, which has been reported in a photoemission study \cite{Veenstra:14}. On this basis, it was shown in an earlier work \cite{Huang:18} that the odd-parity $E_u$ pairing of \SRO~is inevitably 3D in nature, and that, more intriguingly, a novel time-reversal invariant (TRI) nematic $p$-wave pairing may be realized under appropriate circumstances. This would provide a natural explanation for the absence of spontaneous surface current. Also argued was, the system favoring nematic p-wave pairing exhibits only a single phase transition under the uniaxial strain along certain direction as the temperature is lowered. In this paper, we discuss more signatures that could help identify the nematic pairing. Most importantly, when the 3D nematic pairing has a comparatively stronger out-of-plane pairing component, the mean-field $T_c$ enhancement induced by an in-plane uniaxial strain may follow a relation similar to that observed experimentally \cite{Watson:18}. These experiments therefore lend stronger support for the nematic superconducting order. We also argue that the nematic and chiral $p$-wave pairings can be distinguished by the discontinuities in the velocity of certain sound modes in ultrasound absorption at the onset of superconductivity. Finally, we also discuss possible implications of the 3D $E_u$ pairing for $\mu$SR experiment under uniaxial strains and NMR spectroscopy. 

{\bf 3D nematic pairing:} Owing to the weak interlayer coupling, the most frequently discussed $E_u$ pairing classified under the $D_{4h}$ point group symmetry \cite{Sigrist:91} of \SRO~assumes only in-plane pairing $(k_x,k_y)\hat{z}$, where $\hat{z}$ indicates the Cooper pair wavefunction in spin space \cite{Vollhardt:90}. The resultant chiral $p$-wave gap function reads $(k_x+ ik_y)\hat{z}$. However, when the nontrivial 3D SOC is accounted for, the most generic $E_u$ superconducting gap consists of both in-plane (IP) and out-of-plane (OP) channels \cite{Sigrist:91}: $(k_x,k_y)\hat{z}$ and $(k_z\hat{x},k_z\hat{y})$ -- each containing degenerate $x$- and $y$-components. To gain an intuitive understanding for the origin of the 3D pairing, it is instructive to first identify the projection of the orbital and spin angular momenta $|L_z, S_z\rangle$ associated with each of the superconducting components: $k_x \hat{z} \equiv \frac{1}{\sqrt{2}} \left( |1, 0\rangle + |-1,0 \rangle \right)$,
$k_y \hat{z} \equiv \frac{1}{\sqrt{2}i} \left( |1, 0\rangle - |-1,0 \rangle \right)$,
$k_z \hat{x} \equiv  \frac{1}{\sqrt{2}} \left( -|0,1\rangle + |0,-1\rangle \right)$,
$k_z \hat{y} \equiv  \frac{1}{\sqrt{2}i} \left( |0,1\rangle + |0,-1\rangle \right)$.
These are states with Cooper pair angular momentum $J=L \oplus S = 1$ and with equal-weight linear superposition of $J_z=\pm 1$. In the presence of SOC, $J$ and $J_z$ are still good quantum numbers, whilst $L_z$ and $S_z$ are not. Hence the IP and OP pairings are inherently entangled.

The corresponding 3D gap function takes the form $\hat{\Delta}_{\bs k} = [{\bs d}_{\bs k}\cdot {\bs \sigma}]i\sigma_y$, where
\begin{eqnarray}
{\bs d}_{\bs k} &=&  {\bs d}_{x\bs k} + {\bs d}_{y\bs k} \nonumber \\
&=& (\phi_{ix}k_x \hat{z}+\phi_{ox} k_z \hat{x}) + (\phi_{iy}k_y \hat{z}+\phi_{oy} k_z \hat{y})  .
\label{eq:gapD}
\end{eqnarray}
Here, the two expressions in the brackets correspond, respectively, to the $x$ and $y$ components of the $E_u$ pairing, ${\bs d}_{\mu \bs k}$ ($\mu=x,y$). Note that we have used $\phi_{i\mu}$ and $\phi_{o\mu}$  to separately designate the IP and OP pairings. This is necessary as the two channels respond differently to external uniaxial strains, as will become clear later. The nematic state is favored when the OP channel becomes moderately strong \cite{Huang:18}, and in this case only the $x$- or $y$-component condenses in the horizontal nematic phase, while both are condensed in the diagonal nematic phase. By contrast, the chiral pairing state sees a phase difference of $\pm \pi/2$ between the two components.

\begin{figure}
\includegraphics[width=7cm]{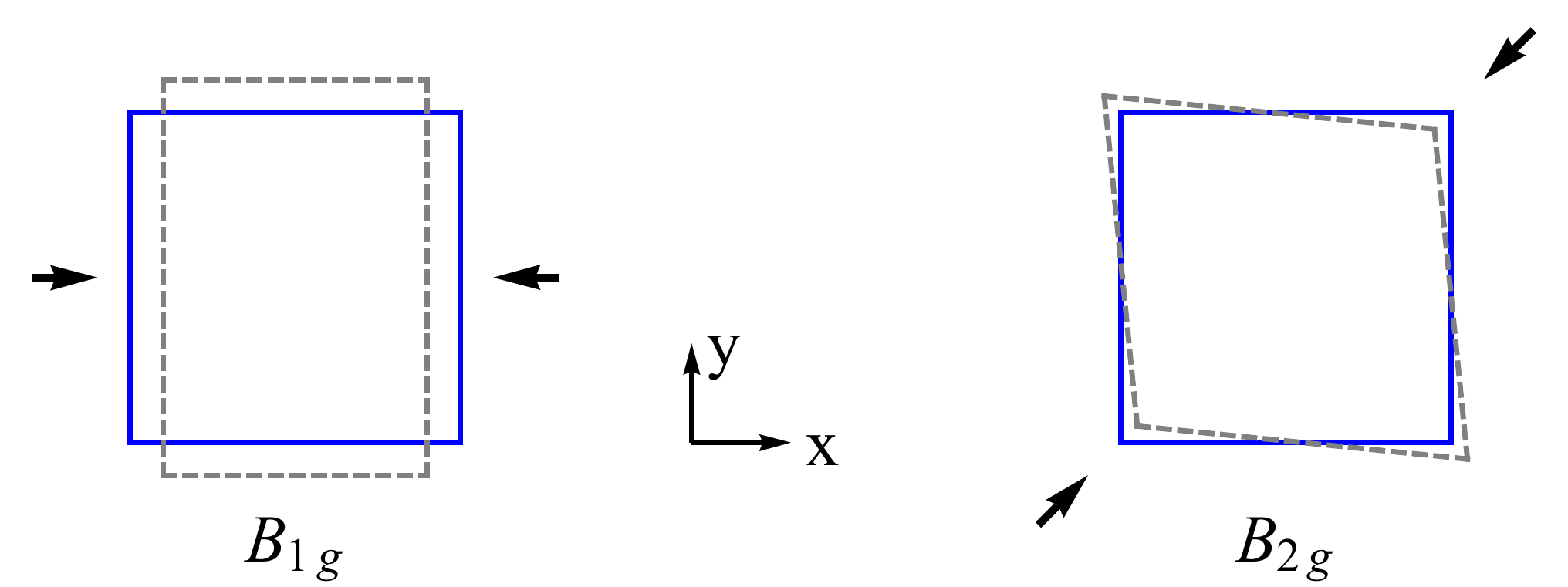}
\caption{Distortion of the lattice structure in \SRO~due to uniaxial $B_{1g}$ and $B_{2g}$ strains. The thick arrows mark the direction of the (compressive) strain applied. The grey dashed polygons illustrate the distorted lattices. The corresponding corrections to the electronic band structure acquire the respective forms of $k_x^2-k_y^2$ and $k_xk_y$. }
\label{fig:strain}
\end{figure}

{\bf Phenomenological theory with strain:} We now proceed to discuss the strain response of the superconducting order parameters and then focus on the scenario where horizontal nematic pairing is favored in the unperturbed system. We first consider an in-plane uniaxial strain with $B_{1g}$ symmetry, as depicted in Fig. \ref{fig:strain}(a). Such a strain is applied parallel to the crystalline $x$- or $y$-axis. When the strain is relative weak, its main effect is to alter the electron dispersion, $\xi_{\bs k} \ra \xi_{\bs k}+\delta\xi_{\bs k}$ with $\delta \xi_{\bs k} \propto \psi (k_x^2-k_y^2)$, where $\psi$ is a dimensionless scalar field measuring the magnitude of the strain and we have set the electron mass to unity for simplicity \cite{Hicks:14}. The focus here is on small-strain response, hence the large-strain limit where the $\gamma$-band may experience a Lifshitz transition \cite{Steppke:17,Liu:17,Barber:18,LuoYK:18} is beyond the scope of our study. 

To investigate the influence of strain on the superconducting order parameters, it suffices to study the Gingzburg-Landau free energy density up to quadratic powers of the superconducting and strain fields,
\begin{eqnarray}
 {\cal F}&=& \alpha_i(T) |\phi_{ix}|^2 + \alpha_o(T) |\phi_{ox}|^2  +  \lambda(\phi_{ix}^\ast\phi_{ox} + c.c.)  \nonumber \\
&+& \alpha_i(T) |\phi_{iy}|^2 + \alpha_o(T) |\phi_{oy}|^2 + \lambda( \phi_{iy}^\ast\phi_{oy} + c.c.) \nonumber \\
&+& a_i \psi (|\phi_{ix}|^2 - |\phi_{iy}|^2 )  + b_i\psi^2 (|\phi_{ix}|^2 + |\phi_{iy}|^2 )   \nonumber \\
&+& b_o \psi^2(|\phi_{ox}|^2 + |\phi_{oy}|^2 ),
\label{eq:fe}
\end{eqnarray}
where the terms with coefficient $\lambda$ describe the coupling between the IP and OP channels of the $E_u$ pairing. As argued in Ref. \onlinecite{Huang:18}, the appearance of these terms can be ascribed to a 3D spin-orbital entanglement. The coefficients $\alpha_{\nu}(T)\approx (T-T_{c\nu})/T_{c0}, (\nu=i,o)$, where $T_{c0}$ denotes the superconducting transition temperature in the absence of strain (evaluated below), and $T_{c\nu}$'s are the respective intrinsic transition temperatures of the IP and OP channels had the SOC been absent. All of the coefficients are dimensionless.

The first and second lines of Eq. (\ref{eq:fe}) describe, respectively, the free energy associated with the two $E_u$ channels. The energy is minimized with a particular composition of the two channels, $\{ \phi_{i\mu},\phi_{o\mu}\} \propto \{ \alpha_i(T)-\alpha_o(T) - \sqrt{[\alpha_i(T)-\alpha_o(T)]^2 + 4\lambda^2},2\lambda \}$. As a consequence, the two shall condense at the same transition $T=T_{c0}$ determined by the relation $\alpha_i(T)+\alpha_o(T) - \sqrt{[\alpha_i(T)-\alpha_o(T)]^2 + 4\lambda^2} =0$. Notably, when the pairing is dominated by either of the two channels, $\lambda$ must be relatively weak compared to the $T_{c\nu}$'s. This is because a sizable $\lambda$ would otherwise lead to considerable mixing between the two channels. It is also noteworthy that the $\lambda$-couplings tend to enhance the overall pairing.

Uniaxial strain influences the two $E_u$ channels in disparate manner. At leading order the IP channel couples linearly to the strain field as $a_i\psi (|\phi_{ix}|^2-|\phi_{iy}|^2)$, while the OP channel does not. In particular, the $B_{1g}$ strain $\psi$ acts as a perturbation that influences `only' (see below) the orbital sector of the Cooper pair wavefunction, i.e. it changes the Cooper pair orbital angular momentum by two while leaving its spin angular momentum invariant, taking the form $\sim |1,0\rangle\langle -1,0| + |-1,0\rangle \langle 1,0|$ where the brackets denote the pairing state with angular momentum $|L_z,S_z\rangle$. It can then be checked that, at linear order the strain perturbs the two IP components separately, and with opposite coupling coefficients as in Eq. (\ref{eq:fe}).

The OP channel is unaffected at this level as their orbital sector is immune to the perturbation at linear order. Nonetheless, due to the finite SOC, the spin sector of the Cooper pair wavefunction in principle also couples to the uniaxial strain at this order. This is because the strain-induced correction to the electronic structure also `propogates' to the spin-orbital entanglement in the electron Bloch wavefunctions. Incidentally, the corrections due to the $B_{1g}$ strains vanish along the Brillouin zone diagonals where the entanglement is strongest. Away from these regions, the modification to the entanglement is down by an additional factor of $(\eta_\text{soc}/E_F)^2 \sim 0.01$. Taken together, it is thus justifiable to neglect the linear-order response of the OP channel, as we shall do throughout this paper.

The quadratic order of the strain $\psi^2$, on the other hand, carries zero angular momentum and hence couples to all superconducting components indiscriminately, as can be seen in Eq. (\ref{eq:fe}). Typically, a concomitant strain component with $A_{1g}$ symmetry must also appear due to the non-unity Poisson's ratio in \SRO~\cite{Hicks:14}. This would introduce linear couplings $\psi_{A_{1g}}(|\phi_{ix}|^2 + |\phi_{iy}|^2)$ and $\psi_{A_{1g}}(|\phi_{ox}|^2 + |\phi_{oy}|^2)$, which merely leads to constant shifts of the superconducting $T_c$ for the individual pairing components \cite{Watson:18}. We hence ignore this contribution in our analyses.

\begin{figure}
\subfigure{\includegraphics[width=4.2cm]{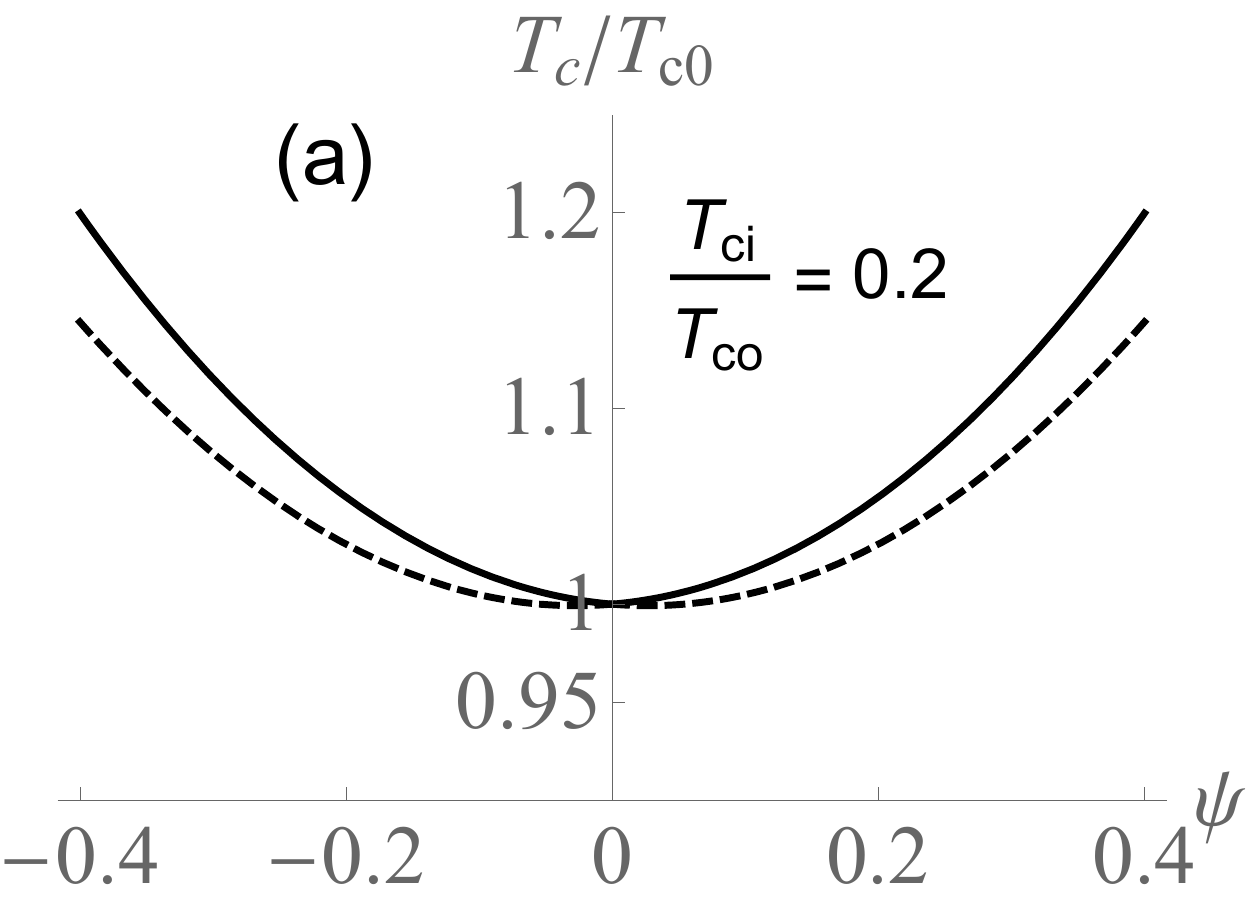} }
\subfigure{\includegraphics[width=4.2cm]{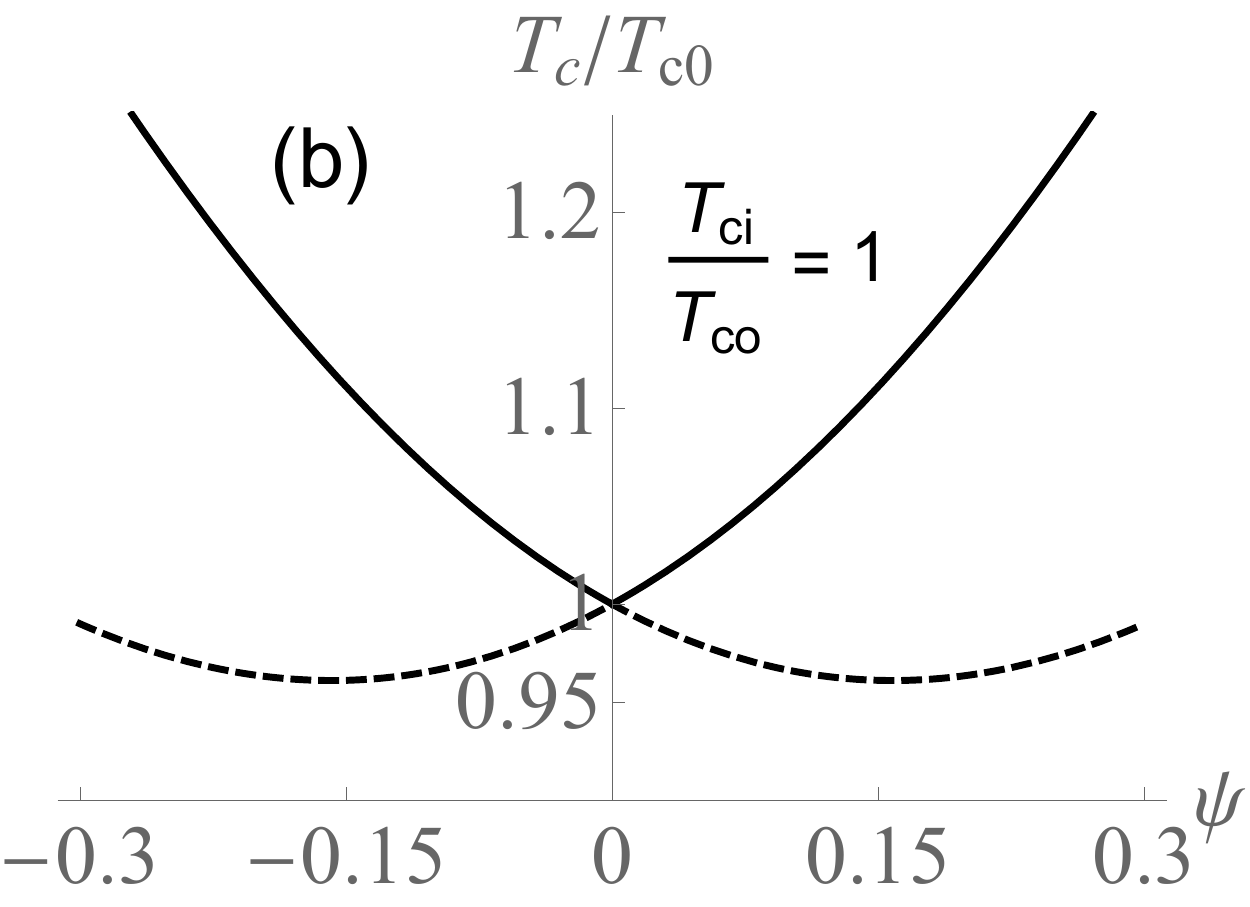} }
\caption{ Splitting of the critical temperatures as a function of the uniaxial $B_{1g}$ strain $\psi$. Solid and dashed curves denote upper and lower $T_c$'s, $T_{c\pm}$, respectively. The ratio $T_{ci}/T_{co}$ determines the relative strength of the IP and OP pairings. Note there exists a very weak cusp in (a) at zero strain. We have chosen the parameters $\alpha_{1,2}=1,\lambda=0.2,a=1,b_{1,2}=-1$. The range of $\psi$ is chosen to roughly reproduce the degree of $T_c$ variation in the recent scanning scope measurement \cite{Watson:18}. Notice that the detailed shape of the curves is sensitive to the parameters. For example, negative $b_{\nu}$'s are typically needed to ensure an enhanced upper critical temperature at larger strains.}
\label{fig:TR}
\end{figure}

If the unperturbed \SRO~stabilizes the horizontal nematic pairing, its response to a finite $B_{1g}$ uniaxial strain may well be consistent the existing observations. For the widely discussed $E_u$ state with vanishing OP pairing, the $T_c$ initially increases linearly over a range of $\psi$ \cite{Sigrist:91}. However, in the opposite limit with purely OP pairing, depending on the sign of $b_o$ the $T_c$ should increase or decrease quadratically as a function of strain $\psi$ at the level of approximation treated in the present study. Normally, $b_o$ is negative such that the strain enhances $T_c$ quadratically with $\psi$. One thus expects that, when the balance tips in favor of the OP channel, the range of linear $\psi$-dependence shrinks, giving way to a predominantly quadratic behavior observed in the recent scanning SQUID measurements \cite{Watson:18}. This is demonstrated in Fig. \ref{fig:TR}, which plots the split critical temperatures of two scenarios, one with dominant OP channel and the other with the two channels comparable. In resonance with our central message, the former shows a weak linear cusp that is hardly discernible, resembling the experimental observation. Note that in this horizontal nematic phase, the dashed lines in Fig. \ref{fig:TR} do not represent true phase transition, as the orthogonal superconducting component never develops at low $T$ \cite{Huang:18}. Hence there exists only one phase transition as marked by the solid line. 

In fact, the weak cusp is expected so long as the OP channel dominates, even if diagonal nematic or chiral pairing is stabilized instead. Both shall observe two transitions under $B_{1g}$ strains. However, compared to the robust lower phase transition breaking time-reversal symmetry in the chiral phase, the lower transition in the diagonal nematic phase -- which breaks a mirror symmetry -- could be smeared in the presence of strain misalignment \cite{Huang:18}. 

Here we caution that the behavior depicted in Fig. \ref{fig:TR} may not be the most representative one. For example, in writing down the free energy Eq. (\ref{eq:fe}), we have ignored possible strain-tuning of the intrinsic $T_{c\nu}$'s. They may well acquire nonlinear dependence on the strain as the $\gamma$-band Fermi surface is driven towards the van-Hove singularity. The van-Hove physics appears to dominate at large applied strains \cite{Hicks:14,Steppke:17}, but whether it plays any significant role at small strains, as in the measurements of Watson {\it et al.} \cite{Watson:18}, remains unclear. Further, the shape of the curves in Fig. \ref{fig:TR} is sensitive to the free energy parameters. Resolving these issues requires a more careful examination of the microscopic model for \SRO.

Similar analyses follow for the strain with $B_{2g}$ symmetry, as illustrated in Fig. \ref{fig:strain}(b). The electronic structure is now corrected by $\delta \xi_{\bs k} \propto \tilde \psi k_xk_y$. In analogy with the previous analyses, the $B_{2g}$ strain represents a perturbation of the following form, $i\left( |1,0\rangle\langle -1,0| - |-1,0\rangle \langle 1,0|\right)$, which also changes the Cooper pair orbital angular momentum by two. At linear order, this strain mixes the two IP pairings but does not affect the OP channel. The quadratic order, ${\tilde \psi}^2$, again couples to all pairing components. The last two lines of the free energy (\ref{eq:fe}) are replaced by,
\begin{eqnarray}
{\cal F}_{\tilde \psi}&=& \tilde a_i \tilde \psi (\phi^\ast_{ix}\phi_{iy} +c.c. )  + \tilde b_i {\tilde \psi}^2 (|\phi_{ix}|^2 + |\phi_{iy}|^2 )   \nonumber \\
&+& \tilde b_o {\tilde \psi}^2 (|\phi_{ox}|^2 + |\phi_{oy}|^2 ) \,.
\end{eqnarray}
Some straightforward algebra reveals qualitatively the same relation for $T_c$. Experimentally, a much flatter strain-dependent $T_c$ was observed \cite{Hicks:14}, possibly due to the much weaker electronic structure distortion induced by the $B_{2g}$ strain \cite{Hicks:14}. This may indeed be the case, given the vanishing of $\delta\xi_{\bs k}$ around the van Hove singular momenta. We shall not elaborate this scenario in details here.

{\bf Ultrasound absorption:} It has been discussed that ultrasound absorption could provide crucial evidence for the chiral $p$-wave pairing \cite{Lupien:01,Lupien:02,Sigrist:02,Fay:00,Kee:00,Higashitani:00,Walker:02}. We shall show that the same measurement could also help to identify the nematic pairings in this material. For convenience, we write the $x$ and $y$ components of the $E_u$ order parameter in Eq. (\ref{eq:gapD}) in a concise form $\bs \Phi = (\Phi_x, \Phi_y)$, as was done in Ref. \onlinecite{Huang:18}. They couple to the lattice vibrations in the following fashion \cite{Sigrist:02},
\begin{eqnarray}
 {\cal F}_\text{sc-el} &=& [r_1 (\epsilon_{xx}+ \epsilon_{yy}) + r_2 \epsilon_{zz}](|\Phi_x|^2 + |\Phi_y|^2) \nonumber \\
&&+r_3 \epsilon_{xy} (\Phi_x^\ast \Phi_y + \Phi_y^\ast \Phi_x) \nonumber \\
&&+r_4(\epsilon_{xx} - \epsilon_{yy})(|\Phi_{x}|^2 - |\Phi_{y}|^2)  \,,
\label{eq:mainGLscel}
\end{eqnarray}
where $\epsilon_{ij}$ represent the basic lattice strain tensor components, and $r_i$ the coupling constants. An $E_u$ pairing possesses a number of low-energy collective modes. This is made obvious by writting $\bs \Phi=\Phi (\cos \theta, e^{i\gamma} \sin \theta)$, where $\Phi, \theta$ and $\gamma$ are all real. Two of the collective modes correspond to the relative phase ($\gamma$) and relative amplitude ($\theta$) fluctuations between the order parameter components. Through (\ref{eq:mainGLscel}), these modes couple to lattice strains and therefore influence the dynamics of certain sound waves below the superconducting transition \cite{footnote1}. In the Supplementary Material \cite{supplementary}, we analyze the renormalization of the shearing elastic constants $c_{66}$ and $c_{11}-c_{12}$ due to the coupling with these collective modes for both chiral and nematic phases. We find that, while both of these constants exhibit discontinuity at the onset of the chiral phase, only one experiences a jump for the nematic phase. Specifically, for the diagonal nematic state,
\begin{equation}
\Delta c_{66} =0, ~\text{and}~\Delta (c_{11}-c_{12}) \propto -\frac{\Delta C_v}{T_c}\left[ \frac{\partial T_c}{\partial (\epsilon_{xx}-\epsilon_{yy})} \right]^2 ,
\label{eq:diagNem}
\end{equation}
while for the horizontal nematic state,
\begin{equation}
\Delta c_{66}  \propto -\frac{\Delta C_v}{T_c}\left( \frac{T_c}{\partial \epsilon_{xy}}\right)^2,~\text{and}~\Delta (c_{11}-c_{12}) =0 \,,
\label{eq:horiNem}
\end{equation}
where $\Delta C_v$ is the specific heat jump at the transition. These shall in turn result in velocity jumps at $T_c$ for the sound modes that are associated with the corresponding shearing elastic constants, thereby providing a means to distinguish the different $E_u$ phases. 

{\bf Discussions and summary:} The nematic odd-parity pairing seems to contradict the signs of TRSB seen experimentally \cite{Luke:98,Xia:06}. Here we propose a possible phenomenon which might reconcile this. In view of the $Z_2$ symmetry breaking associated with the doubly degenerate nematic states, domains of different nematic orientations could coexist, as has indeed been implicated in a number of measurements \cite{Kidwingira:06,Saitoh:15,Wang:17,Anwar:17}. For a state breaking $U(1)\times Z_2$ symmetry, $Z_2$ domains can be formed, the corners of which could attach $Z_2$ vortices carrying half quantum fluxes \cite{Sigrist:99,Isacsson:05,Xu:07}. Further, when domain walls traverse lattice defects, dipole-like magnetic fluxes may be pinned, as has been numerically demonstrated in a different context \cite{Garaud:14,Garaud:16}. Thus, the TRSB in the bulk of \SRO~as seen by $\mu$SR \cite{Luke:98} may be explained. Nonetheless, it remains unclear how a nonvanishing Kerr rotation \cite{Xia:06} could be explained, although a recent proposal of odd-frequency pairing may be noted \cite{Komendova:17}. 

A remark is in order about when the system is subject to uniaxial strains. A $B_{1g}$ strain, for example, lifts the aforementioned double-degeneracy of the horizontal nematic but not that of the diagonal nematic pairing. Hence, the latter still permits nematic domains and $Z_2$ vortices below a lower phase transition, and, in principle, likewise for the former under a $B_{2g}$ strain. Similary, $\mu$SR can probe the internal magnetic field so generated.  

Further, since a 3D $E_u$ pairing, chiral or nematic, contains both in-plane ($\hat{x}$ or $\hat{y}$) and out-of-plane ($\hat{z}$) ${\bs d}$-vector components, the spin susceptibility under generic magnetic field orientations shall be suppressed below the superconducting transition. In the nematic state, the degree of suppression depends on the magnetic field orientation. A horizontal nematic pairing with only finite ${\bs d}_{x\bs k}$, for instance, shall see a drop in the Knight shift under any magnetic field except for $\bs B \parallel \bs y$. We note that a substantial reduction has indeed been observed in a recent NMR Knight shift measurement \cite{Luo:19}. 

Despite the promise to explain some outstanding puzzles, the nematic pairing is yet to be checked against other notable observations. For example, the indications of line-nodal gap structure \cite{Nishizaki:00,Hassinger:16,Kittaka:18} is inconsistent with the point-nodes of a simple nematic state. However, the highly anisotropic electronic structure could in principle generate line nodes unrelated to symmetry \cite{Huang:18,Roising:18,Acharya:18}, or simply deep gap minima \cite{Wang:18,Firmo:13,Dodaro:18}. Nonetheless, more measurements, such as those proposed in Ref. \onlinecite{Huang:18}, shall be examined.

Favoring the nematic pairing requires a dominant out-of-plane pairing, for which a satisfactory physical justification is still lacking. Noteworthily, in the weak-coupling approach reminiscent of the celebrated Kohn-Luttinger mechanism for Cooper instabilities \cite{Kohn:65,Shankar:94,Raghu:10a}, the predicted pairing symmetry and gap structure in \SRO~could become highly sensitive to microscopic details adopted in a particular calculation \cite{Scaffidi:14,Tsuchiizu:15,Zhang:17b,Liu:17,Wang:18}. Thus, in light of its nontrivial 3D spin-orbital entanglement, a 3D generalization of the microscopic calculations would be an interesting direction to further pursue.

In summary, we demonstrated how the proposed 3D nematic odd-parity pairing could be reconciled with a number of perplexing observations that defy a definitive chiral $p$-wave interpretation. Besides the absence of spontaneous surface current \cite{Kirtley:07,Hicks:10,Curran:14}, these also include measurements performed on samples subject to symmetry-lowering perturbations, such as in-plane uniaxial strains and magnetic fields \cite{Hicks:14,Steppke:17,Watson:18,Yonezawa:14}. We showed that, when the out-of-plane pairing dominates, the superconducting $T_c$ enhanced by the strain may follow a relation resembling a quadratic behavior, and the linear cusp may be too weak to resolve experimentally. As a side remark, both superconducting fluctuations \cite{Fischer:16} and strain impurities \cite{YuY:19} could in principle operate in parallel to further smear the linear cusp even in the case of weaker out-of-plane pairing. Further, while the chiral phase of the $E_u$ pairing shall observe two robust phase transitions under the stated symmetry-breaking perturbations, the nematic phase may exhibit a single transition \cite{Huang:18}. We finally argued how nematic domain proliferation may explain the internal magnetic signal seen by $\mu$SR \cite{Luke:98} and discussed a possible identification of the nematic pairing through ultrasound measurements. 

{\bf Acknowledgement:} We would like to thank Clifford Hicks, Manfred Sigrist, Qiang-Hua Wang, Chris Watson, Fan Yang, and Fu-Chun Zhang for valuable discussions. This work is supported in part by the NSFC under grants No. 11825404 (W.H. and H.Y.) and No.11774306 (Y.Z.), by the National Key Research and Development Program of China under grant No.2016YFA0300202 (Y.Z.) and No. 2016YFA0301001 (H.Y.), by the Strategic Priority Research Program of Chinese Academy of Sciences under Grant No. XDB28000000 (Y.Z. and H.Y.), and by the C.N. Yang Junior Fellowship at Tsinghua University (W.H.).

\widetext
\begin{center}
\textbf{\large Supplemental Material for ``Possible 3D nematic odd-parity pairing in \SRO: experimental evidences and predictions"}\\
\vspace{4mm}
\author{Wen Huang}
\affiliation{Institute for Advanced Study, Tsinghua University, Beijing 100084, China}
\author{Yi Zhou}
\email{yizhou@zju.edu.cn}
\affiliation{Department of Physics, Zhejiang University, Hangzhou, Zhejiang 310027, China}
\affiliation{CAS Center for Excellence in Topological Quantum Computation, University of Chinese Academy of Sciences, Beijing 100190, China}
\author{Hong Yao}
\email{yaohong@tsinghua.edu.cn}
\affiliation{Institute for Advanced Study, Tsinghua University, Beijing 100084, China}
\affiliation{ State Key Laboratory of Low Dimensional Quantum Physics, Tsinghua University, Beijing 100084, China }
\end{center}

\section{Ultrasound Absorption in the $E_u$ superconducting state}
\subsection{A. Formulism} 
We consider a generic two-component superconducting order parameter in the $E_u$ channel under the $D_{4h}$ group, $\bs\Phi=(\Phi_x,\Phi_y)$. Following Sigrist [64], and dropping the gradient contributions, the generic Ginzburg-Landau free energy density takes the form,
\begin{equation}
f_\text{sc} = \alpha (|\Phi_x|^2 + |\Phi_y|^2) + \beta_1 (|\Phi_x|^2+|\Phi_y|^2)^2 + \frac{\beta_{2}}{2}[(\Phi_x^\ast \Phi_y)^2 + (\Phi_y^\ast\Phi_x)^2]+  \beta_3|\Phi_x|^2 |\Phi_y|^2
\label{eq:GLsc}
\end{equation}
where $\alpha= \alpha^\prime (T-T_c)$, and the elastic energy density is
\begin{eqnarray}
f_\text{el} &=& \frac{1}{2}[c_{11}(\epsilon_{xx}^2 + \epsilon_{yy}^2) + c_{33} \epsilon_{zz}^2 + 2c_{12}\epsilon_{xx}\epsilon_{yy} + 4c_{66}\epsilon_{xy}^2 + 2c_{13}(\epsilon_{xx}+\epsilon_{yy})\epsilon_{zz}+4c_{44}(\epsilon_{xz}^2+\epsilon_{yz}^2)] \nonumber \\
&=& \frac{1}{2}[(c_{11}-c_{12})(\epsilon_{xx}- \epsilon_{yy})^2+c_{12}(\epsilon_{xx}^2 + \epsilon_{yy}^2) + 2c_{11}\epsilon_{xx}\epsilon_{yy}+ c_{33} \epsilon_{zz}^2  + 4c_{66}\epsilon_{xy}^2 + 2c_{13}(\epsilon_{xx}+\epsilon_{yy})\epsilon_{zz}+4c_{44}(\epsilon_{xz}^2+\epsilon_{yz}^2)] \nonumber \\
&&
\label{eq:GLel}
\end{eqnarray}
where $\epsilon_{ij}$ are components of the lattice strain tensor. Finally, the coupling between strain and the superconducting order parameters reads,
\begin{equation}
f_\text{sc-el} = [r_1 (\epsilon_{xx}+ \epsilon_{yy}) + r_2 \epsilon_{zz}](|\Phi_x|^2 + |\Phi_y|^2) + r_3 \epsilon_{xy} (\Phi_x^\ast \Phi_y + \Phi_y^\ast \Phi_x) + r_4(\epsilon_{xx} - \epsilon_{yy})(|\Phi_{x}|^2 - |\Phi_{y}|^2)
\label{eq:GLscel}
\end{equation}
Expressing the order parameter as $\bs \Phi=\Phi (\cos \theta, e^{i\gamma} \sin \theta)$ where $\Phi, \theta$ and $\gamma$ are all real, the possible superconducting states are characterized by different equilibrium values of $(\theta,\gamma) = (\theta_0,\gamma_0)$, i.e. $(\pi/4,\pm \pi/2)$ for the chiral phase, $(\pi/4,0)$ for the diagonal nematic state, and $(0,0)$ for the horizontal nematic state. Collective modes, which are associated with fluctuations from $\theta_0$ and $\gamma_0$, couple to certain strain tensor components and renormalize the corresponding elastic constant $\epsilon_{ij}$. It is our goal here to identify the elastic constants that are influenced by the collective fluctuations. To proceed, we rewrite the free energy terms that involve the superconducting order parameters,
\begin{equation}
f^\prime = a \Phi^2 + \beta_1 \Phi^4 + \frac{\Phi^4}{4} (\beta_2 \cos 2\gamma +\beta_3) \sin^2(2\theta)  + r_1 (\epsilon_{xx}+\epsilon_{yy}) \Phi^2 + r_3 \epsilon_{xy} \Phi^2 \sin 2\theta \cos \gamma + r_4 (\epsilon_{xx}-\epsilon_{yy}) \Phi^2\cos 2\theta
\label{eq:GLnew}
\end{equation}
We then expand this free energy in terms of the small deviations from the unperturbed state: $\theta^\prime$ and $\gamma^\prime$ (we neglect the trivial $\Phi$-fluctuation for simplicity). After a straightforward saddle point approximation with respect to $\theta^\prime$ and $\gamma^\prime$, we obtain the renormalization to the elastic moduli. Similar analysis has been performed for the chiral state by Sigrist [64], while we repeat the procedure below for generality.

\subsection{B. Chiral state}
It can be seen from Eq. (\ref{eq:GLnew}) that to stabilize a chiral state with $(\theta_0,\gamma_0)=(\pi/4,\pi/2)$, the system must satisfy the relation $\beta_2>0$ and $-\beta_2+\beta_3<0$. Substituting $(\theta,\gamma)= (\theta_0+\theta^\prime,\gamma_0+\gamma^\prime)$ into Eq. (\ref{eq:GLnew}),

\begin{eqnarray}
f^\prime &=&  \frac{\Phi^4}{4} [-\beta_2 \cos 2\gamma^\prime+\beta_3  ] \cos^22\theta^\prime - r_3 \epsilon_{xy}\Phi^2 \cos 2\theta^\prime \sin \gamma^\prime - r_4 (\epsilon_{xx}-\epsilon_{yy})\Phi^2\sin 2\theta^\prime + ... ...  \nonumber \\
&=&\frac{\beta_2 \Phi^4 }{2} (\gamma^\prime)^2 - r_3 \epsilon_{xy} \Phi^2 \gamma^\prime  +  (\beta_2-\beta_3) \Phi^4  (\theta^\prime)^2 - 2r_4(\epsilon_{xx}-\epsilon_{yy}) \Phi^2 \theta^\prime  +  \mathcal{O}[(\theta^\prime)^2, (\gamma^\prime)^2, \theta^\prime\gamma^\prime]
\end{eqnarray}
Minimizing with respect to $\theta^\prime$ and $\gamma^\prime$, one obtains,
\begin{equation}
f^\prime \approx -\frac{ r_3^2 \epsilon_{xy}^2}{2\beta_2} -\frac{ r_4^2(\epsilon_{xx} - \epsilon_{yy})^2}{\beta_2-\beta_3}
\end{equation}
Comparing with Eq. (\ref{eq:GLel}), we obtain the renormalization of the elastic constant $c_{66}$ and $c_{11} -c_{12}$ induced by the relative phase and amplitude modes, respectively,
\begin{eqnarray}
\Delta c_{66} &=& -\frac{r_3^2 }{4 \beta_2} \\
\Delta(c_{11}-c_{12}) &=& -\frac{2r^2_4}{\beta_2-\beta_3}
\label{eq:Cre}
\end{eqnarray}
To relate this to the specific heat jump $\Delta C_v$ at $T_c$, we recognize the following relations,
\begin{eqnarray}
&&\frac{\Delta C_v}{T_c} =\frac{2 (a^\prime)^2}{4\beta_1-\beta_2 +\beta_3}  \\
&&\frac{\partial T_c}{ \partial \epsilon_{xy}}=-\frac{r_3}{a^\prime} \\
&&\frac{\partial T_c}{ \partial (\epsilon_{xx} - \epsilon_{yy})} = -\frac{r_4}{a^\prime}
\end{eqnarray}
Combining these with Eq. (\ref{eq:Cre}), we obtain,
\begin{eqnarray}
\Delta c_{66} &=& - \frac{\Delta C_v}{T_c}  \left(\frac{\partial T_c}{\partial \epsilon_{xy}} \right)^2 \frac{4\beta_1-\beta_2+\beta_3}{8\beta_2} \\
\Delta(c_{11}-c_{12}) &=&   - \frac{\Delta C_v}{T_c}  \left[\frac{\partial T_c}{\partial (\epsilon_{xx}-\epsilon_{yy})} \right]^2 \frac{4\beta_1-\beta_2+\beta_3}{\beta_2-\beta_3}
\end{eqnarray}
which are consistent with Sigrist where they apply.

\subsection{C. Diagonal nematic state}
In the nematic state with $(\theta_0,\gamma_0)=(\pi/4,0)$, the system has the relation $\beta_2, \beta_2+\beta_3<0$. Writting in terms of $\theta^\prime$ and $\gamma^\prime$ Eq. (\ref{eq:GLnew}) becomes,

\begin{eqnarray}
f^\prime &=&  \frac{\Phi^4}{4} [ \beta_2 \cos 2\gamma^\prime+\beta_3  ] \cos^2 2\theta^\prime + r_3 \epsilon_{xy}\Phi^2 \cos 2\theta^\prime \cos \gamma^\prime - r_4 (\epsilon_{xx}-\epsilon_{yy})\Phi^2\sin 2\theta^\prime + ... ...  \nonumber \\
&=& - \frac{\beta_2 \Phi^4 }{2} (\gamma^\prime)^2 - \frac{r_3 \epsilon_{xy} \Phi^2}{2} (\gamma^\prime)^2 - 2 r_3 \epsilon_{xy} \Phi^2 (\theta^\prime)^2  - (\beta_2+\beta_3) \Phi^4  (\theta^\prime)^2 - 2r_4(\epsilon_{xx}-\epsilon_{yy}) \Phi^2 \theta^\prime  +  \mathcal{O}[(\theta^\prime)^2, (\gamma^\prime)^2, \theta^\prime\gamma^\prime]  \nonumber \\
~~
\end{eqnarray}
One can see that the relative phase fluctuation $\gamma^\prime$ is not influenced by the strain components.  We thus minimize $f^\prime$ with respect to $\theta^\prime$, obtaining
\begin{equation}
f^\prime \approx \frac{ r_4^2(\epsilon_{xx} - \epsilon_{yy})^2}{\beta_2+ \beta_3+ \frac{2r_3\epsilon_{xy}}{\Phi^2}}
 \approx    \frac{ r_4^2(\epsilon_{xx} - \epsilon_{yy})^2}{\beta_2+ \beta_3}  \end{equation}
Comparing with Eq. (\ref{eq:GLel}), we obtain the renormalization of the elastic constant $c_{11}-c_{12}$ induced by the relative amplitude mode,
\begin{eqnarray}
\Delta(c_{11}-c_{12}) &=& \frac{2r^2_4}{\beta_2+\beta_3}  =   \frac{\Delta C_v}{T_c}  \left[\frac{\partial T_c}{\partial (\epsilon_{xx}-\epsilon_{yy})} \right]^2 \frac{4\beta_1-\beta_2+\beta_3}{\beta_2+\beta_3}
\end{eqnarray}

\subsection{D. Horizontal nematic state}
This state has $(\theta_0,\gamma_0)=(0,0)$, and the coefficients satisfy $\beta_2<0 < \beta_2+\beta_3$. Expressed in terms of $\theta^\prime$ and $\gamma^\prime$ Eq. (\ref{eq:GLnew}) becomes,

\begin{eqnarray}
f^\prime &=&  [ (\beta_2+\beta_3)\Phi^4 -2r_4(\epsilon_{xx}-\epsilon_{yy})\Phi^2] (\theta^\prime)^2 + 2r_3 \epsilon_{xy} \Phi^2 \theta^\prime +  \mathcal{O}[(\theta^\prime)^2, (\gamma^\prime)^2, \theta^\prime\gamma^\prime]
\end{eqnarray}
Once again the relative phase fluctuation $\gamma^\prime$ is not influenced by the strain components at this order. Minizing with respect to $\theta^\prime$ leads to,
\begin{equation}
f^\prime \approx -\frac{ r_3^2 \epsilon_{xy}^2}{\beta_2+ \beta_3- \frac{2r_4(\epsilon_{xx}-\epsilon_{yy})}{\Phi^2}}
 \approx     -\frac{ r_3^2 \epsilon_{xy}^2}{\beta_2+ \beta_3 }
\end{equation}
We then reach the renormalization of the elastic constant $c_{66}$ induced by the relative apmlitude fluctuations,
\begin{eqnarray}
\Delta c_{66} &=& -\frac{2 r^2_3}{\beta_2+\beta_3}   = - \frac{\Delta C_v}{T_c}  \left(\frac{\partial T_c}{\partial \epsilon_{xy}} \right)^2 \frac{4\beta_1-\beta_2+\beta_3}{\beta_2+\beta_3}
\end{eqnarray}

\subsection{E. Conclusion}
In the chiral phase, the strain components $\epsilon_{xy}$ and $\epsilon_{xx}-\epsilon_{yy}$ couple respectively to the relative phase and relative amplitude fluctuations, leading to discontinuities at $T_c$ in the shearing elastic constant $c_{66}$ and $c_{11}-c_{12}$. In the diagonal nematic state, there is a discontinuity in $c_{11}-c_{12}$, but not in $c_{66}$, vice versa for the horizontal nematic state. Notably, in both nematic phases, it is the relative amplitude mode which couples to the elastic constant exhibiting discontinuity at the superconducting transition.


\begin{thebibliography}{99}
\bibitem{footnote1} Note we have checked that the additional collective fluctuations between the IP and OP channels do not enter the coupling in Eq. \ref{eq:mainGLscel}.
\end{thebibliography}
\end{document}